\documentclass[12pt]{article}
\usepackage{pst-plot,epsf}
\usepackage{axodraw}
\newcommand\pubnumber{DESY 01-029\\LC-TH-2001-035}
\newcommand\pubdate{March 7, 2001}
\newcommand\hepnumber{~}

\def\csumb{Deutsches Elektronen-Synchrotron DESY \\
Platanenallee 6, \\ D--15738 Zeuthen, Germany}

\def\Title#1{\begin{center} {\Large\bf #1 } \end{center}}
\def\Author#1{\begin{center}{ \sc #1} \end{center}}
\def\Address#1{\begin{center}{ \it #1} \end{center}}

\newcommand\pubblock{\rightline{\begin{tabular}{l} \pubnumber\\
         \pubdate\\ \hepnumber \end{tabular}}}
\newenvironment{Abstract}{\begin{quotation}  }{\end{quotation}}

\makeatletter
\def\section{\@startsection{section}{0}{\z@}{5.5ex plus .5ex minus
 1.5ex}{2.3ex plus .2ex}{\large\bf}}
\def\subsection{\@startsection{subsection}{1}{\z@}{3.5ex plus .5ex minus
 1.5ex}{1.3ex plus .2ex}{\normalsize\bf}}
\def\subsubsection{\@startsection{subsubsection}{2}{\z@}{-3.5ex plus
-1ex minus  -.2ex}{2.3ex plus .2ex}{\normalsize\sl}}

\renewcommand{\@makecaption}[2]{%
   \vskip 10pt
   \setbox\@tempboxa\hbox{\small #1: #2}
   \ifdim \wd\@tempboxa >\hsize     
       \small #1: #2\par          
     \else                        
       \hbox to\hsize{\hfil\box\@tempboxa\hfil}
   \fi}

 \def\citenum#1{{\def\@cite##1##2{##1}\cite{#1}}}
\def\citea#1{\@cite{#1}{}}
 
\newcount\@tempcntc
\def\@citex[#1]#2{\if@filesw\immediate\write\@auxout{\string\citation{#2}}\fi
  \@tempcnta\z@\@tempcntb\m@ne\def\@citea{}\@cite{\@for\@citeb:=#2\do
    {\@ifundefined
       {b@\@citeb}{\@citeo\@tempcntb\m@ne\@citea\def\@citea{,}{\bf ?}\@warning
       {Citation `\@citeb' on page \thepage \space undefined}}%
    {\setbox\z@\hbox{\global\@tempcntc0\csname b@\@citeb\endcsname\relax}%
     \ifnum\@tempcntc=\z@ \@citeo\@tempcntb\m@ne
       \@citea\def\@citea{,}\hbox{\csname b@\@citeb\endcsname}%
     \else
      \advance\@tempcntb\@ne
      \ifnum\@tempcntb=\@tempcntc
      \else\advance\@tempcntb\m@ne\@citeo
      \@tempcnta\@tempcntc\@tempcntb\@tempcntc\fi\fi}}\@citeo}{#1}}
\def\@citeo{\ifnum\@tempcnta>\@tempcntb\else\@citea\def\@citea{,}%
  \ifnum\@tempcnta=\@tempcntb\the\@tempcnta\else
  {\advance\@tempcnta\@ne\ifnum\@tempcnta=\@tempcntb \else\def\@citea{--}\fi
    \advance\@tempcnta\m@ne\the\@tempcnta\@citea\the\@tempcntb}\fi\fi}
\makeatother

\def\NPB{{\em Nucl. Phys.} B }

\def\PLB{{\em Phys. Lett.}  B }

\def\PRD{{\em Phys. Rev.} D }
\def\ZPC{{\em Z. Phys.} C }

\def\amu{a_\mu}

\def\MZ{M_Z}

\def\az{\alpha(\MZ)}
\def\aiz{\alpha^{-1}(\MZ)}
\def\dalf{\Delta\alpha}
\def\das{\Delta\alpha(s)}

\def\dah{\Delta\alpha^{(5)}_{\rm had}}
\def\dahs{\Delta\alpha^{(5)}_{\rm had}(s)}
\def\dahz{\Delta\alpha^{(5)}_{\rm had}(\MZ^2)}
\def\dahzE{\Delta\alpha^{(5)}_{\rm had}(-\MZ^2)}
\def\dah0{\Delta\alpha^{(5)}_{\rm had}(-s_0)}

\newcommand{\gv}{\mbox{GeV}}

\newcommand{\MOM}{${\mathrm{MOM}}$ }

\newcommand{\MSb}{$\overline{\mathrm{MS}}$ }

\newcommand{\MSbm}{\overline{\mathrm{MS}} }
\newcommand{\al }{\alpha}
\newcommand{\epm}{e^+e^-}

\newcommand{\be}{\begin{equation}}
\newcommand{\ee}{\end{equation}}
\newcommand{\ba}{\begin{eqnarray}}
\newcommand{\ea}{\end{eqnarray}}
\newcommand{\bea}{\begin{eqnarray*}}
\newcommand{\eea}{\end{eqnarray*}}
\newcommand{\bet}{\begin{center} \begin{tabular}}
\newcommand{\ent}{\end{tabular} \end{center}}
\newcommand{\bb}{\begin{thebibliography}}
\newcommand{\eb}{\end{thebibliography}}

\newcommand{\ra}{\rightarrow}

\newcommand{\bit}{\begin{itemize}}
\newcommand{\eit}{\end{itemize}}

\newcommand{\veps}{\varepsilon}

\newcommand{\lapprox}{\raisebox{-.2ex}{$\stackrel{\textstyle<}
{\raisebox{-.6ex}[0ex][0ex]{$\sim$}}$}}

\newcommand{\crn}{\nonumber \\}
\newcommand{\noi}{\noindent}
\newcommand{\nn}{\nonumber}
\newcommand{\ha}{\frac{1}{2}}
\newcommand{\dal}{\Delta \alpha}

\newcommand{\sha}{\sigma(e^+e^- \rightarrow {\rm hadrons})}
\newcommand{\mz}{M^2_Z}
\newcommand{\mw}{M^2_W}
\newcommand{\de}{\frac{\delta e}{e}}

\newcommand{\sinf}{\sin^2 \Theta_f}
\newcommand{\cosf}{\cos^2 \Theta_f}
\newcommand{\sini}{\sin^2 \Theta_i}
\newcommand{\cosi}{\cos^2 \Theta_i}
\newcommand{\sinW}{\sin^2 \Theta_W}
\newcommand{\cosW}{\cos^2 \Theta_W}
\newcommand{\sing}{\sin^2 \Theta_g}

\newcommand{\dro}{\Delta \rho}
\newcommand{\Gmu}{G_{\mu}}
\newcommand{\bary}{\begin{array}}
\newcommand{\eary}{\end{array}}

\newcommand{\stt}{s_\theta^2}
\newcommand{\ctt}{c_\theta^2}
\newcommand{\ctf}{c_\theta^4}
\newcommand{\cts}{c_\theta^6}
\newcommand{\ctit}{c_\theta^{-2}}

\begin{document}
\begin{titlepage}
\pubblock

\vfill
\def\thefootnote{\fnsymbol{footnote}}
\Title{The Effective Fine Structure Constant at TESLA Energies}
\vfill
\Author{F. Jegerlehner}
\Address{\csumb}
\vfill
\begin{Abstract}
We present a new estimate of the hadronic contribution to the shift in
the fine structure constant at LEP and TESLA energies and calculate
the effective fine structure constant. Substantial progress in a
precise determination of this important parameter is a consequence of
substantially improved total cross section measurements by the BES II
collaboration and an improved theoretical understanding.  In the
standard approach which relies to a large extend on experimental data
we find $\Delta \al _{\rm hadrons}^{(5)}(\mz) = 0.027896 \pm 0.000395$
which yields $\alpha^{-1}(\mz) = 128.907 \pm 0.054$.  A few values at
higher energies are given in the following table:
\begin{center}
\begin{tabular}{|r|c|c|}
\hline
\hline
$\sqrt{s}$ GeV  & $\dahs$ & $\alpha^{-1}(s)$\\
\hline
 100 &0.0283 $\pm$ 0.0004 & 128.790 $\pm$ 0.054 \\
 300 &0.0338 $\pm$ 0.0004 & 127.334 $\pm$ 0.054 \\
 500 &0.0372 $\pm$ 0.0004 & 126.543 $\pm$ 0.054 \\
 800 &0.0417 $\pm$ 0.0004 & 125.634 $\pm$ 0.054 \\
1000 &0.0436 $\pm$ 0.0004 & 125.229 $\pm$ 0.054 \\
\hline
\end{tabular}
\end{center}
Another approach, using the Adler function as a tool to compare theory
and experiment, allows us to to extend the applicability of
perturbative QCD in a controlled manner. The result in this case reads
$\Delta\alpha^{(5)}_{\rm had}(M_Z^2) = 0.027730 \pm 0.000209$ and
hence $\alpha^{-1}(\mz) = 128.930 \pm 0.029$. At TESLA energies a new
problem shows up with the definition of an effective charge. A
possible solution of the problem is presented.  Prospects for further
progress in a precise determination of the effective fine structure
constant are discussed.
\end{Abstract}
\vfill
\end{titlepage}
\def\thefootnote{\arabic{footnote}}
\setcounter{footnote}{0}

\parindent 0mm
\parskip 2mm
\renewcommand{\arraystretch}{1.4}

\section{Introduction}

Precision physics requires appropriate inclusion of higher order
effects and the knowledge of very precise input parameters of the
electroweak standard model SM. One of the basic input parameters is
the fine structure constant which depends logarithmically on the
energy scale. Vacuum polarisation effects lead to a partial screening
of the charge in the low energy limit (Thomson limit) while at higher
energies the strength of the electromagnetic interaction grows. In
this note we have in mind future precision physics at TESLA~\cite{TDR}
energies as a continuation of LEP~\cite{LEP} experiments and thus
consider the effective fine structure constant at energies up to 1
TeV. Very likely, TESLA in addition of being a gauge boson factory
like LEP will be a Higgs factory.\\

Renormalization of the electric charge $e$ by a shift $\delta
e$ at different scales leads to a shift of the fine structure
constant by
\be
\Delta \alpha = 2 \left( \de (0) - \de (M_Z)\right)
= \Pi'_{\gamma}(0)-\Pi'_{\gamma}(\mz) 
\ee
where $\Pi'_{\gamma}(s)$ is the photon vacuum polarisation function
defined via the time-ordered product of two electromagnetic currents
$j_{em}^\mu(x)$:
\be
i\int d^4x\,e^{iq\cdot x}\langle 0| {\rm T} j_{em}^\mu(x) j_{em}^\nu(0)
                        |0 \rangle
= -(q^2 g^{\mu\nu} - q^\mu q^\nu)\Pi'_{\gamma}(q^2)\;.
\ee
The shift $\Delta \alpha$ is large due to the large change in scale
going from zero momentum to the Z-mass scale $\mu=M_Z$ and due to the
many species of fermions contributing. Zero momentum more precisely
means the light fermion mass thresholds.

In perturbation theory the leading light fermion ($m_f \ll M_Z$)
contribution is given by

\hfill \epsfbox{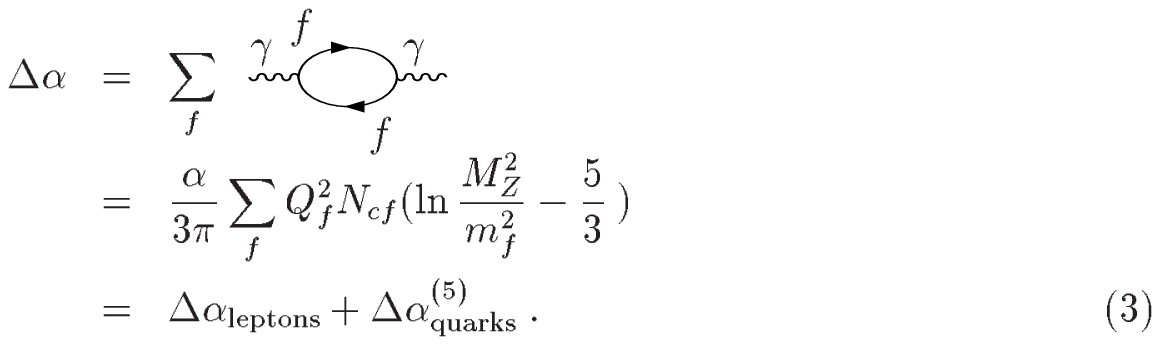}

\setcounter{equation}{3}
\label{dallep}

A serious problem is the low energy contributions of the five light
quarks u,d,s,c and b which cannot be reliably calculated using
perturbative quantum chromodynamics (p-QCD). The evaluation of the
hadronic contribution $\Delta \al _{\rm quarks}^{(5)} \ra \Delta \al
_{\rm hadrons}^{(5)}$ is the main concern of this note. Before I am
going into this, let me make a few remarks about its consequences for
precision physics.

A major drawback of the partially non-perturbative relationship
between $\alpha(0)$ and $\alpha(M_Z)$ is that one has to rely on
experimental data exhibiting systematic and statistical errors which
implies a non-negligible uncertainty in our knowledge of the
effective fine structure constant. In precision predictions of gauge
boson properties this has become a limiting factor.  Since $\alpha~,~
G_\mu, M_Z$ are the most precisely measured parameters, they are used
as input parameters for accurate predictions of observables like the
effective weak mixing parameter $\sinf$, the vector $v_f$ and
axial-vector $a_f$ neutral current couplings, the $W$ mass $M_W$ the
widths $\Gamma_Z$ and $\Gamma_W$ of the $Z$ and the $W$, respectively,
etc. However, for physics at higher energies we have to use the
effective couplings at the appropriate scale, for physics at the
$Z$--resonance, for example, $\alpha(M_Z)$ is more adequate to use
than $\alpha(0)$. Of course this just means that part of the higher
order corrections may be absorbed into an effective parameter. If we
compare the precision of the basic parameters
\be \bary{ccccccccccc}
\frac{\delta \alpha}{\alpha} &\sim& 3.6 &\times& 10^{-9}&~~~~~&
\frac{\delta \alpha(M_Z)}{\alpha(M_Z)} &\sim& 1.6 \div 6.8 &\times& 10^{-4}\\
\frac{\delta G_\mu}{G_\mu} &\sim& 8.6 &\times& 10^{-6}&&
\frac{\delta M_Z}{M_Z} &\sim& 2.4 &\times& 10^{-5}
\eary
\ee
we observe that the uncertainty in $\alpha(M_Z)$ is roughly an order
of magnitude worse than the next best, which is the $Z$--mass. Let me
remind the reader that $\dal$ enters in electroweak precision physics
typically when calculating versions of the weak mixing parameter
$\sini$ from $\al$, $\Gmu$ and $M_Z$ via
\be
\sini\:\cosi\: =\frac{\pi \al}{\sqrt{2}\:G_\mu\:M_Z^2} 
\frac{1}{1-\Delta r_i}
\ee
where 
\ba
\Delta r_i &=&\Delta r_i({\al ,\: \Gmu ,\: M_Z ,}\:m_H,
\:{m_{f\neq t},\:m_t})
\ea
includes the higher order corrections which can be calculated in the
SM or in alternative models. It has been calculated for the first time
by A.~Sirlin in 1980~\cite{Sirlin80}. In the SM the Higgs mass $m_H$ is the
only relevant unknown parameter and by confronting the calculated with
the experimentally determined value of $\sini$ one obtains the
important indirects constraints on the Higgs mass. $\Delta r_i$
depends on the definition of $\sini$. The various definitions coincide
at tree level and hence only differ by quantum effects. From the weak
gauge boson masses, the electroweak gauge couplings and the neutral
current couplings of the charged fermions we obtain
\ba
\sinW &=& 1-\frac{M_W^2}{M_Z^2}\\
\sing &=& e^2/g^2=\frac{\pi \al}{\sqrt{2}\:G_\mu\:M_W^2}\\
\sinf &=& 
\frac{1}{4|Q_f|}\;\left(1-\frac{v_f}{a_f} \right)\;,\;\;f\neq \nu\;,
\ea
for the most important cases and the general form of $\Delta r_i$ reads
\ba
\Delta r_i &=& \dal - f_i(\sini)\:\dro + \Delta r_{i\:\mathrm{remainder}}
\label{der}
\ea
with a universal term $\dal$ which affects the predictions for $M_W$,
$A_{LR}$, $A^f_{FB}$, $\Gamma_f$, etc. The oder terms can be
calculated safely in perturbation theory. $\dro$ is the famous
correction to the $\rho$--parameter, first calculated by M.~Veltman in
1977~\cite{Veltman77}\footnote{The article for the first time
established non--decoupling heavy particle effects in spontaneously
broken gauge theories and presented the calculation of heavy fermion
contributions. While $\dro$ measures the the weak iso-spin breaking
proportional to the sum of $|m_{t'}^2-m_{b'}^2|$, where $t'$ and $b'$
denote the top and bottom components of weak iso--doublets and thus
vanishes for mass degenerate doublets,the latter contribute to another
parameter $\Delta_1$ which is then given by $\Delta_1=\frac{\alpha}{24
\pi s_W^2}\:N_d$ and directly measures the number $N_d$ of degenerate heavy
doublets. $\dro$ and $\Delta_1$ nowadays go under the labels $\veps_1$
or $\alpha T$ and $\veps_3$ or $\frac{\alpha S}{4s_W^2}$, respectively
($s_W^2 \equiv \sinW$, $c_W^2=1-s_W^2$).}, exhibiting the leading top
mass correction
\ba
\dro \simeq \frac{\sqrt2 \Gmu}{16 \pi^2}\:3m_t^2\;\;;\;\;\;m_t \gg m_b
\ea 
which allowed LEP experiments to obtain a rather good indirect
estimate of the top quark mass prior to the discovery at the
TEVATRON~\cite{TopTev95}. Note that in~(\ref{der}) $f_W=c_W^2/s_W^2
\simeq 3.35$ is substantially enhanced relative to $f_f=1$. The
``remainder'' term although sub-leading is very important for the
interpretation of the precision experiments at LEP and includes part
of the leading Higgs mass dependence. For a heavy Higgs particle we
obtain the simple expression
\ba
\Delta r^{\rm Higgs}_i \simeq \frac{\sqrt2 \Gmu \mw}{16 \pi^2}\:
\left\{c^H_i\: (\ln \frac{m_H^2}{\mw}- \frac56) \right\}
\;\;;\;\;\;m_M \gg M_W
\ea
where $c^H_f=(1+9 \sinf)/(3\sinf)$ and $c^H_W=11/3$, for example.

The uncertainty $\delta \Delta \alpha$ implies uncertainties $\delta
M_W$, $\delta \sinf$ given by
\ba
\frac{\delta M_W}{M_W} &\sim& \ha \frac{\sinW}{\cosW-\sinW}
\;\delta \dal \sim 0.23 \;\delta \dal \\
\frac{\delta \sinf}{\sinf} &\sim& ~~\frac{\cosf}{\cosf-\sinf}
\;\delta \dal \sim 1.54 \;\delta \dal\;
\ea
which obscure in particular the indirect bounds on the Higgs mass
obtained from electroweak precision measurements. A summary of the
present status and future expectations will be presented below. Once
the Higgs boson will have been discovered and its mass is known, precision
measurements of the $\Delta r_i$, which would be possible with the
Giga-Z option of TESLA~\cite{TDR}, would provide excellent possibilities
to establish new physics contributions beyond the SM.
Similar tests would be possible by confronting the effective parameters
\ba
\hat{G}_\mu = \frac{12\pi \Gamma_{W \ell \nu}}{\sqrt2 M_W^3}
\;\;{\rm and}\;\;\;
\hat{\rho}  = \frac{M_W^3}{M_Z^3} 
\frac{2\Gamma_{Z\nu\nu}}{\Gamma_{W \ell \nu}}
\ea
which are the high energy versions of $\Gmu$ and $\rho \equiv G_{\rm
NC}(0)/\Gmu$ which are not plagued by uncertainties from $\dal$. Here,
$G_{\rm NC}(0)$ denotes the low energy effective neutral current
coupling.

\section{The hadronic contributions to $\alpha(s)$}

The effective QED coupling constant at scale $\sqrt{s}$ is given by
the renormalization group resummed running fine structure constant
\be
\alpha(s) =  \frac{\alpha}{1 - \das}
\ee
with
\be
\das =- 4\pi\alpha
        {\rm Re}\left[ \Pi'_{\gamma}(s) - \Pi'_{\gamma}(0)
                \right]\;.
\ee

\vspace*{46mm}

\begin{figure}[h]
\begin{picture}(120,60)(-20,0)
{\scalebox{.75 .75}{%
\epsfbox{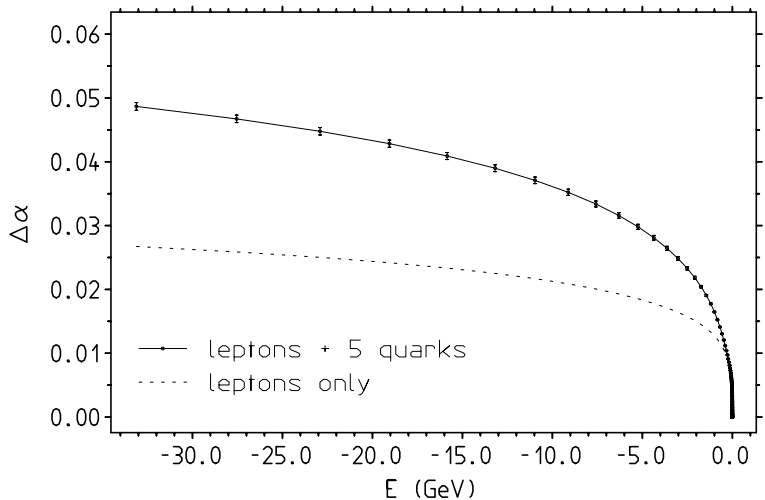}}}
\end{picture}

\caption[]{The running of $\alpha$. The ``negative'' $E$ axis is
chosen to indicate space-like momentum transfer. The vertical bars at
selected points indicate the uncertainty.}
\label{fig:alpharun}
\end{figure}
Figure~\ref{fig:alpharun} shows the running of $\alpha$ at low
space-like momentum transfer. The leptonic contributions are
calculable in perturbation theory where at leading order the free
lepton loops yield
\be \bary{l}
\dalf_{{\rm leptons}}(s)= \cr \bary{lcl}
& = & \sum\limits_{\ell=e,\mu,\tau}
      \frac{\alpha}{3\pi}
      \left[ - \frac{8}{3} + \beta_\ell^2
             - \frac{1}{2}\beta_\ell(3 - \beta_\ell^2)
               \ln\left( \frac{1-\beta_\ell}{1+\beta_\ell}
                  \right)
      \right]   \cr
& = &
      \sum\limits_{\ell=e,\mu,\tau}
      \frac{\alpha}{3\pi}
      \left[ \ln\left( s/m_\ell^2
                \right)
           - \frac{5}{3}
           + O\left( m_\ell^2/s
              \right)
      \right]  {\rm \ for \ } |s|\gg m_\ell^2   \cr
& \simeq & 0.03142 {\rm \ for \ } s=M_Z^2 \eary \eary
\ee
where $\beta_\ell = \sqrt{1 - 4m_\ell^2/s}$. This leading contribution
is affected by small electromagnetic corrections only in the next to
leading order. The leptonic contribution is actually known to three
loops~\cite{KalSab55,Ste98} at which it takes the value
\be 
\Delta \alpha_{\rm leptons} (M_Z^2) \; \simeq \; 314.98 \: \times \: 10^{-4}. 
\ee 

In contrast the corresponding free quark loop contribution gets
substantially modified be low energy strong interaction effects, which
cannot be obtained by p-QCD.
Fortunately one can evaluate this hadronic term $\Delta \al
_{\rm hadrons}^{(5)}$ from hadronic $\epm $- annihilation data by using a
dispersion relation. The relevant vacuum polarisation amplitude
satisfies the convergent dispersion relation
\bea
Re \Pi'_{\gamma}(s)-\Pi'_{\gamma}(0)=\frac{s}{\pi} Re \int_{s_0}^{\infty}
ds' \frac{Im \Pi'_{\gamma}(s')}{s'(s'-s -i \veps)} \nn
\eea
and using the optical theorem (unitarity) one has
\bea
Im \Pi'_{\gamma}(s)=\frac{s}{e^2}
                    \sigma_{tot} (\epm \ra \gamma^* \ra {\rm hadrons})(s)
\;. \nn
\eea
In terms of the cross-section ratio
\bea
R(s)=\frac{\sigma_{tot} (\epm \ra \gamma^* \ra {\rm hadrons})}
     {\sigma (\epm \ra \gamma^* \ra \mu^+ \mu^-)}\;, \nn
\eea
where $\sigma (\epm \ra \gamma^* \ra \mu^+ \mu^-)=\frac{4\pi
\al ^2}{3s}$ at tree level, we finally obtain
\bea
\Delta \al _{\rm hadrons}^{(5)}(M_Z^2) &=&
-\frac{\al M_Z^2}{3\pi}Re\int_{4m_{\pi}^2}^{\infty}
ds\frac{R(s)}{s(s-M_Z^2-i\veps)}\;.
\eea
Using the experimental data for $R(s)$ up to $\sqrt{s}=E_{cut}=5.5$
GeV and for the $\Upsilon$ resonances region between 9.6 and 11 GeV
and perturbative QCD from 5.5 to 9.6 GeV and for the high energy
tail~\cite{GKL,ChK95,ChHK00} above 11 GeV we get as an update
of~\cite{EJ95} including the recent new data from CMD~\cite{CMD} and
BES~\cite{BES}
\bea
\Delta \al _{\rm hadrons}^{(5)}(\mz) &=& 0.027896 \pm 0.000395 \\
\alpha^{-1}(\mz) &=& 128.907 \pm 0.054\;\;.  
\eea

\vspace*{48mm}

\begin{figure}[h]
\begin{picture}(120,60)(-20,0)
{\scalebox{.86 .86}{%
\epsfbox{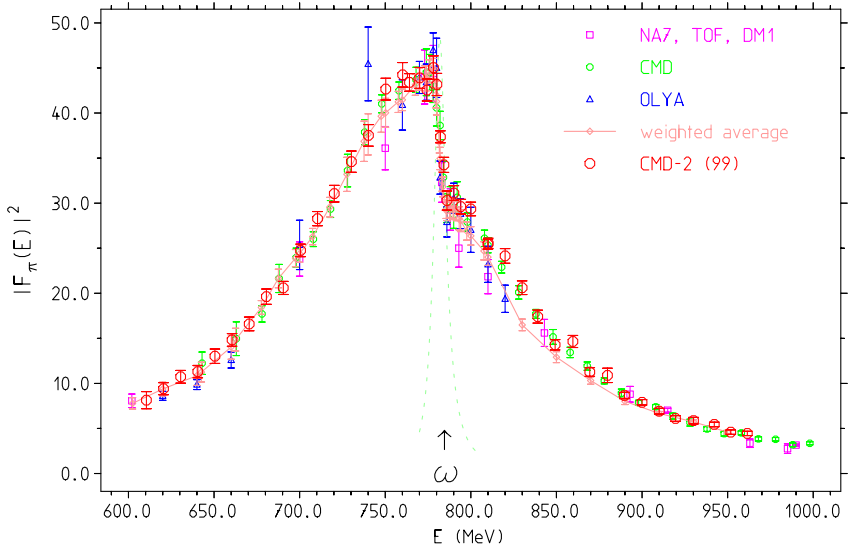}}}
\end{picture}

\caption[]{Recent CMD-2 results~\cite{CMD}.}
\label{fig:CMD2}
\end{figure}
at $M_Z=$ 91.19 GeV. The CMD-2 experiment at Novosibirsk has continued
and substantially improved the $\sha$ measurements below 1.4
GeV~\cite{CMD} and the BES-II experiment at Beijing has published a
new measurement, which in the region from 2 to 5 GeV improves the
evaluation from 15\% to 20\% systematic error to about
6.6\%~\cite{BES}. As a consequence we observe a dramatic reduction of
the error with respect to our 1995 evaluation $0.0280\pm
0.0007$~\cite{EJ95} mainly due to the new BES data. The latter result
has been independently confirmed earlier in~\cite{BP95,ADH98}. For an
evaluation which yields a quite different answer see~\cite{MOR00} and
Tab.~\ref{tab:alperr} below.

From the BES-II data we have subtracted the narrow resonance
$\Upsilon(3)$ (6 points) because this resonance contribution is
calculated as a Breit--Wigner resonance using the parameters of the
particle data table~\cite{PDG}. The BES-II data in the $J/psi$
resonance range from 3.6 GeV to 5 GeV is integrated and combined with
results for integral obtained from other experiments. Below the
resonance we calculate the weighted average with the older data points
and integrate the weighted average. This procedure has been motivated
and tested in~\cite{EJ95}.

\begin{figure}[h]
\begin{picture}(120,60)(-20,0)
{\scalebox{.86 .86}{%
\epsfbox{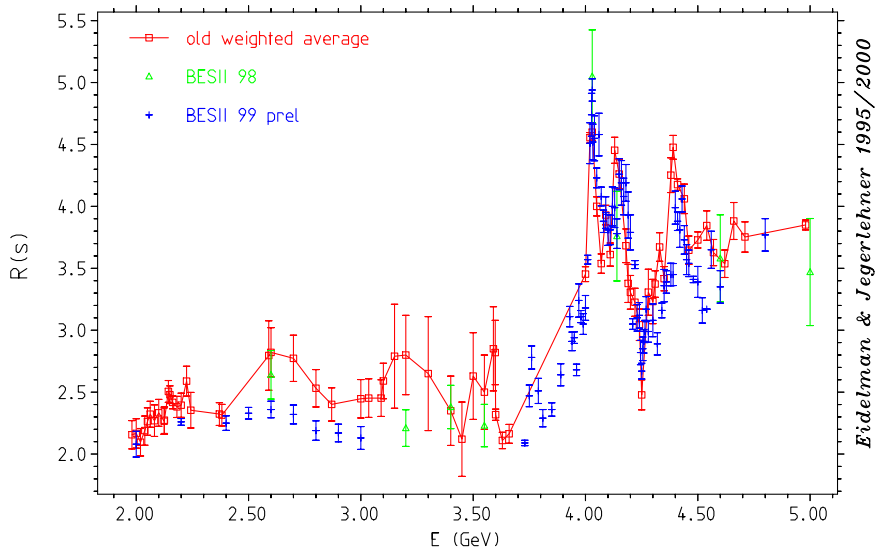}}}
\end{picture}

\caption[]{Recent BES-II results~\cite{BES}.}
\label{fig:BES2}
\end{figure}

For details about our evaluation procedure of we refer to~\cite{EJ95}.
In our {\em standard approach} we take data serious as published and
combine them according to rules suggested by the particle data
group~\cite{PDG}. In Figs.~\ref{fig:CMD2} and \ref{fig:BES2} we show
the new data in comparison to the older ones.

Below we will present another result obtained with the {\em Euclidean
approach}, which is based on comparing experimental data and theory
(i.e. p--QCD) by means of the Adler function.

\vspace*{55mm}

The contributions from different energy ranges are shown in Tab.~\ref{tab:dal}

\begin{table}[h]
\begin{tabular}{cc||r||r}
\hline
 final state &  energy range (GeV) & $\dahz$ (stat) (syst)
& $\dah0$ (stat) (syst) \\
\hline
$\chi PT$  &    (0.28, 0.32) &    0.03 ( 0.00) ( 0.00)&      0.03 ( 0.00) ( 0.00) \\
$\rho$     &    (0.28, 0.81) &   25.67 ( 0.21) ( 0.42)&     23.81 ( 0.20) ( 0.39) \\
$\omega$   &    (0.42, 0.81) &    2.96 ( 0.04) ( 0.08)&      2.70 ( 0.03) ( 0.07) \\
$\phi$     &    (1.00, 1.04) &    5.14 ( 0.07) ( 0.12)&      4.41 ( 0.06) ( 0.10) \\
$J/\psi$   &                 &   11.90 ( 0.58) ( 0.64)&      4.21 ( 0.20) ( 0.20) \\
$\Upsilon$ &                 &    1.24 ( 0.05) ( 0.07)&      0.07 ( 0.00) ( 0.00) \\
    hadrons&    (0.81, 1.40) &   13.92 ( 0.11) ( 0.71)&     11.91 ( 0.09) ( 0.59) \\
    hadrons&    (1.40, 3.10) &   26.75 ( 0.10) ( 1.83)&     15.56 ( 0.06) ( 1.13) \\
    hadrons&    (3.10, 3.60) &    5.26 ( 0.05) ( 0.17)&      1.89 ( 0.02) ( 0.06) \\
    hadrons&    (3.60, 9.46) &   50.73 ( 0.24) ( 2.97)&      8.34 ( 0.04) ( 0.44) \\
    hadrons&    (9.46,12.00) &   13.47 ( 0.16) ( 1.19)&      0.72 ( 0.01) ( 0.06) \\
    perturb& (12.0,$\infty$) &  121.67 ( 0.00) ( 0.12)&      1.27 ( 0.00) ( 0.00) \\
\hline			                                                          
    data   &    (0.28,12.00) &  157.05 ( 0.70) ( 3.84)&     73.61 ( 0.31) ( 1.43) \\
    total  &                 &  278.72 ( 0.70) ( 3.84)&     74.89 ( 0.31) ( 1.43) \\
\hline
\end{tabular}
\caption{%
Distribution of uncertainties for $\dahz^{\mathrm{data}}$ and for
$\dah0^{\mathrm{data}}$ in comparison to $\dahz^{\mathrm{data}}$
($\sqrt{s_0}=2.5$ GeV).}
\label{tab:dal}
\end{table}

Note that the $\rho$ contribution slightly increased due to the new
CMD data. The new BES data imply a small increase of the contribution
from the range (3.6, 5.0) GeV, while, as is obvious from
Fig.~\ref{fig:BES2}, the contribution in the range (2.0,3.6) is lower
as compared to previous results. In total the shift in the central
value is very moderate, while the uncertainly has become smaller.

\section{Theoretical progress}

In view of the increasing precision LEP experiments have achieved
during the last few years, more accurate theoretical prediction became
desirable. As elaborated in the introduction, one of the limiting
factors is the hadronic uncertainty of $\dahz$. Because of the large
uncertainties in the data, many authors advocated to extend the use of
perturbative QCD in place of
data~\cite{DH98a,KS98,GKNS98,DH98b,Erler98}. The assumption that p-QCD
may be reliable to calculate
\be
R_\gamma(s) \equiv \frac{\sigma(e^+e^- \rightarrow \gamma^*
\rightarrow {\rm hadrons})}{ \sigma(e^+e^- \rightarrow \gamma^* \rightarrow
\mu^+ \mu^-)} = 12\pi{\rm Im}\Pi'_{\gamma}(s)
\label{RS}
\ee
down to energies as low as 1.8 GeV seems to be supported by
\begin{itemize}
\item the apparent applicability of p-QCD to $\tau$ physics. In fact the
running of $\alpha_s(M_\tau) \ra \alpha_s(M_Z)$ from the $\tau$ mass
up to LEP energies agrees well with the LEP value. The estimated
uncertainty may be debated, however.
\item the smallness~\cite{DH98a} (see also:~\cite{FJ86}) of
non--perturbative (NP) effects if parametrized as prescribed by the
operator product expansion (OPE) of the electromagnetic current
correlator~\cite{SVZ}
\ba
\label{NP}
\Pi_\gamma^{'\mathrm{NP}}(Q^2) &=& \frac{4\pi\al}{3} \sum\limits_{q=u,d,s}
Q_q^2 N_{cq}\,
\cdot \bigg[\frac{1}{12}
\left(1-\frac{11}{18}a\right)
\frac{<\frac{\alpha_s}{\pi} G G>}{Q^4} \crn
&+&2\,\left(1+\frac{a}{3} +\left(\frac{11}{2}-\frac34
l_{q\mu}
\right)\,a^2\right)
\frac{<m_q \bar{q}q>}{Q^4} \\
&+& \left(\frac{4}{27}a
+\left(\frac{4}{3}\zeta_3-\frac{257}{486}-\frac13 l_{q\mu}
\right)\,
a^2\right)
\sum\limits_{q'=u,d,s} \frac{<m_{q'} \bar{{q'}}{q'}>}{Q^4}\,\bigg] \crn
&+& \cdots
 \nn
\ea
where $a\equiv \alpha_s(\mu^2)/\pi$
and $l_{q\mu}\equiv\ln(Q^2/\mu^2)$. $<\frac{\alpha_s}{\pi} G G>$ and
$<m_q \bar{q}q>$ are the scale-invariantly defined condensates.
\end{itemize}

Progress in p-QCD comes mainly from~\cite{mqcd3}. In addition an exact
two--loop calculation of the renormalization group (RG) in the
background field \MOM scheme (BF-MOM) is available~\cite{JT98}. This
allows us to treat ``threshold effects'' closer to physics than in the
\MSb scheme. The BF-MOM scheme respects the QCD Slavonv-Taylor
identities (non-Abelian gauge symmetry) but in spite of that is gauge
parameter ($\xi$) dependent\footnote{In applications considered below
all numerical calculations have been performed in the ``Landau gauge''
$\xi=0$.}. Except from Ref.~\cite{KS98} which is based on~\cite{mqcd3}
most other ``improved'' calculations utilize older results, mainly,
the well known massless result~\cite{GKL} plus some leading mass
corrections. For a recent critical review of the newer estimates of
vacuum polarization effects see~\cite{FJ98} and Tab.\ref{tab:alperr}
below.

\begin{figure}[h]
\begin{picture}(120,60)(-20,0)
\rput{90}(8,4.5){\scalebox{.6 .6}{%
\epsfbox{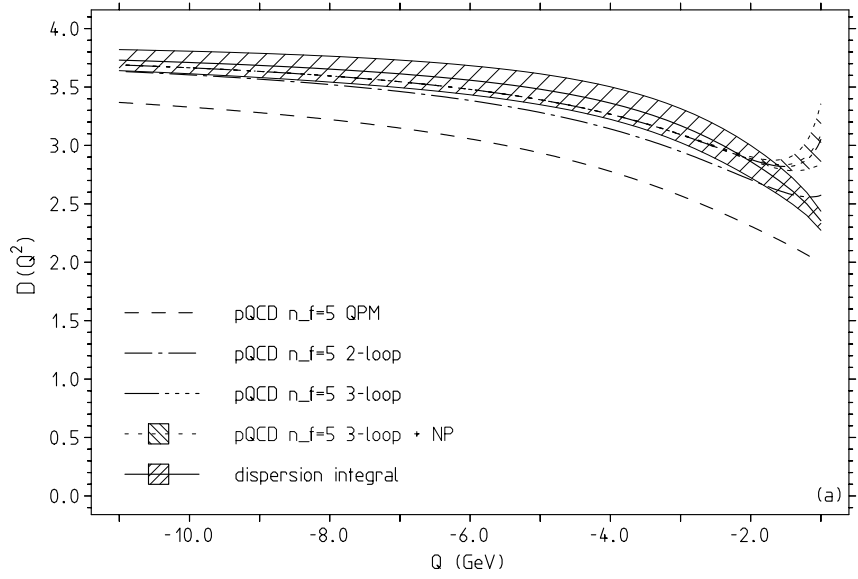}}}
\end{picture}

\caption[]{Adler function: theory vs. experiment (a)~\cite{EJKV98}.}
\label{fig:Adlera}
\end{figure}

\begin{figure}[h]
\begin{picture}(120,60)(-20,0)
\rput{90}(8,4.5){\scalebox{.6 .6}{%
\epsfbox{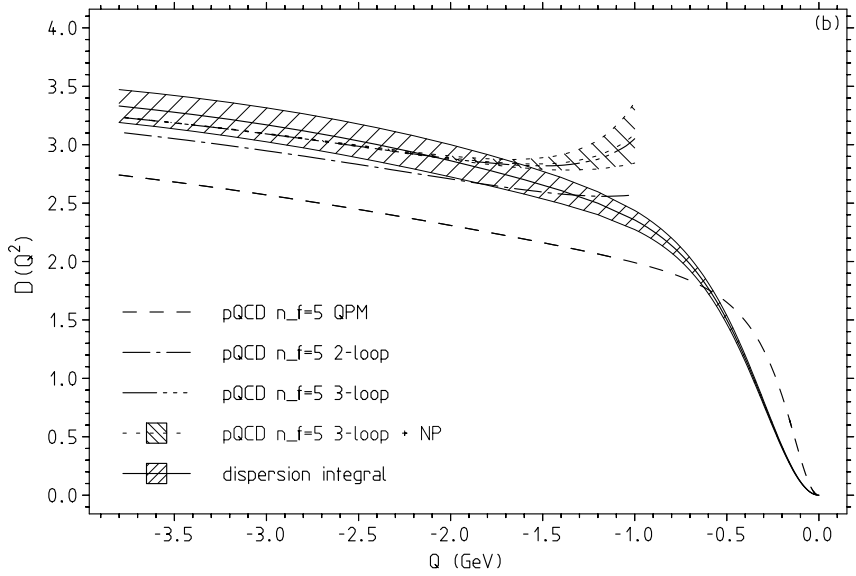}}}
\end{picture}

\caption[]{Adler function: theory vs. experiment (b)~\cite{EJKV98}.}
\label{fig:Adlerb}
\end{figure}

In Ref.~\cite{EJKV98} a different approach of p-QCD improvement was
proposed, which relies on the fact that the vacuum polarization
amplitude $\Pi(q^2)$ is an analytic function in $q^2$ with a cut in
the $s$--channel $q^2=s \geq 0$ at $s\geq 4m^2_\pi$ and a smooth
behavior in the $t$--channel (space-like or Euclidean region). Thus,
instead of trying to calculate the complicated function $R(s)$, which
obviously exhibits non-perturbative features like resonances, one
considers the simpler Adler function in the Euclidean
region. In~\cite{EJKV98} the Adler function was investigated and p-QCD
was found to work very well above 2.5 GeV, provided the exact
three--loop mass dependence was used (in conjunction with the
background field \MOM scheme). The Adler function may be defined as a
derivative
\be
D(-s)=-(12\pi^2)\,s\,\frac{d\Pi'_{\gamma}\,(s)}{ds}
=\frac{3\pi}{\alpha} s\frac{d}{ds}\Delta \alpha_{\mathrm{had}}(s)
\label{DD}
\ee
of (\ref{RS}) which is the hadronic contribution to the shift of the
fine structure constant. It is represented by
\ba
D(Q^2)=Q^2\:\left(\int_{4 m_{\pi}^2}^{E^2_{\rm cut}}
\frac{  R^{\rm data}(s)}{(s+Q^2)^2}ds\;+\;
\int_{E^2_{\rm cut}}^{\infty}\frac{  R^{\rm pQCD}(s)}{(s+Q^2)^2}ds\:\right)
\label{DI}
\ea
in terms of the experimental $\epm$--data. The standard evaluation
(\cite{EJ95}) of (\ref{DI})  then yields the non--perturbative
``experimental'' Adler function, as displayed in
Figs.~\ref{fig:Adlera} and ~\ref{fig:Adlerb}.

For the p-QCD evaluation it is mandatory to utilize the calculations
with massive quarks which are available up to
three--loops~\cite{mqcd3}.  The four-loop corrections are known in the
approximation of massless quarks~\cite{GKL}. The outcome of this
analysis is pretty surprising and is shown in Figs.~\ref{fig:Adlera}
and ~\ref{fig:Adlerb}. For a discussion we refer to the original
paper~\cite{EJKV98}.


\vspace*{56mm}

\vspace*{52mm}

According to (\ref{DD}), we may compute the hadronic vacuum
polarization contribution to the shift in the fine structure constant
by integrating the Adler function. In the region where p-QCD works
fine we integrate the p-QCD prediction, in place of the data. We thus
calculate in the Euclidean region
\be
\Delta\alpha^{(5)}_{\rm had}(-M_Z^2)
=\left[\Delta\alpha^{(5)}_{\rm had}(-M_Z^2) -\Delta\alpha^{(5)}_{\rm
had}(-s_0)\right]^{\mathrm{p-QCD}}+ \Delta\alpha^{(5)}_{\rm
had}(-s_0)^{\mathrm{data}}\;\;.
\ee
A save choice is $s_0=(2.5\, \gv)^2$ where we obtain
\ba
\Delta\alpha^{(5)}_{\rm had}(-s_0)^{\mathrm{data}} =0.007489 \pm
0.000146
\ea
from the evaluation of the dispersion integral (\ref{RS}).
With the results presented above we find
\ba
\Delta\alpha^{(5)}_{\rm
had}(-M_Z^2) = 0.027685 \pm 0.000146 \pm 0.000149[0.000101] 
\ea
for the Euclidean ($t$--channel) effective fine structure constant.
The second error comes from the variation of the pQCD parameters. In
square brackets the error if we assume the uncertainties from
different parameters to be uncorrelated.  The uncertainties coming
from individual parameters are listed in the following table (masses
are the pole masses):

\begin{center}
\begin{tabular}{cccc}
\hline
\hline
parameter & range & pQCD uncertainty & total error \\
\hline
$\alpha_s$ & 0.117 ... 0.123 & 0.000051 & 0.000155 \\ 
$m_c$      & 1.550 ... 1.750 & 0.000087 & 0.000170 \\
$m_b$      & 4.600 ... 4.800 & 0.000011 & 0.000146 \\
$m_t$      & 170.0 ... 180.0 & 0.000000 & 0.000146 \\
\multicolumn{2}{c}{all correlated} & 0.000149 & 0.000209 \\
\multicolumn{2}{c}{all uncorrelated} & 0.000101 & 0.000178 \\
\hline
\end{tabular}
\end{center}

The largest uncertainty is due to the poor knowledge of the charm
mass.  I have taken errors to be 100\% correlated. The uncorrelated
error is also given in the table.

Remaining problems are the following:\\ \noi {\bf a)} contributions to
the Adler function up to three--loops all have the same sign and are
substantial. Four-- and higher--orders could still add up to
non-negligible contribution. An error for missing higher order terms
is not included. The scheme dependence \MSb versus background field
\MOM has been discussed in Ref.~\cite{JT98}.\\ \noi {\bf b)} 
The effective fine structure constant in the time--like region
($s$--channel), as required for $\epm$--collider physics may be
obtained from the Euclidean one by adding the difference
\ba
\Delta=\dahz-\Delta\alpha^{(5)}_{\rm had}(-M_Z^2) =0.000045 \pm
0.000002\,,
\label{delta}
\ea
which may be calculated perturbatively or directly from the
``non--perturbative''\footnote{Since we utilize p-QCD for the high
energy tail in the dispersion integral, $\Delta (s)$ for large $s$ is
dominated by the tail and thus in fact is perturbative.} dispersion
integral. It accounts for the $i\pi$--terms $$\ln (-q^2/\mu^2) =
\ln(|q^2/\mu^2|)+i\pi$$ from the logs.\\ \noi {\bf c)} One may ask the
question whether these terms should be resummed at all, i.e., included
in the running coupling.  Usually such terms tend to cancel against
constant rational terms which are not included in the renormalization
group (RG) evolution. It should be stressed that the Dyson summation
(propagator bubble summation) in general is not a systematic
resummation of leading, sub-leading etc. terms as the RG resummation
is.

It is worthwhile to stress here that the running coupling is {\bf not}
a true function of $q^2$ (or even an analytic function of $q^2$) but a
function of the RG scale $\mu^2$. The coupling as it appears in the
Lagrangian in any case must be a constant, albeit a $\mu^2$--dependent
one, if we do not want to end up in conflict with basic principles of
quantum field theory.  The effective identification of $\mu^2$ with a
particular value of $q^2$ must be understood as a subtraction
(reference) point.\\ The above result was obtained using the
background--field \MOM renormalization scheme, mentioned before. In
the transition from the \MSb to the \MOM scheme we adapt the rescaling
procedure described in~\cite{JT98}, such that for large $\mu$
\bea
\overline{\alpha}_s((x_0\mu)^2)=\alpha_s(\mu^2)+0+ O(\alpha_s^3)\;.
\eea
This means that $x_0$ is chosen such that the couplings coincide
to leading and next--to--leading order at asymptotically large
scales. Numerically we find $x_0\simeq 2.0144$. Due to this
normalization by rescaling the coefficients of the Adler--function
remain the same in both schemes up to three--loops. In the \MOM scheme
we automatically have the correct mass dependence of full QCD, i.e.,
we have automatic decoupling and do not need decoupling by hand and matching
conditions like in the \MSb scheme. For the numerical
evaluation we use the pole quark masses~\cite{PDG}
$m_c=1.55 \gv ,\;m_b=4.70 \gv,
\;m_t=173.80 \gv \;$ and the strong interaction coupling
$\alpha_{s\; {\MSbm}}^{(5)}(M_Z) = 0.120 \pm 0.003$.
For further details we refer to~\cite{EJKV98}.

Since $\Delta$ Eq.~(\ref{delta}) is small we may include it in the resummation
without further worrying and thus obtain
\ba
\Delta\alpha^{(5)}_{\rm
had}(M_Z^2) = 0.027730 \pm 0.000209[0.000178] \crn 
\alpha^{-1}(\mz) = 128.930 \pm 0.029[0.024]
\;\;.
\ea
The alternative evaluation by the Euclidean approach is
compared with the standard evaluation in Tab.~\ref{tab:dal}. 

\begin{figure}[hb]
\begin{picture}(120,60)(30,20)
\epsfbox{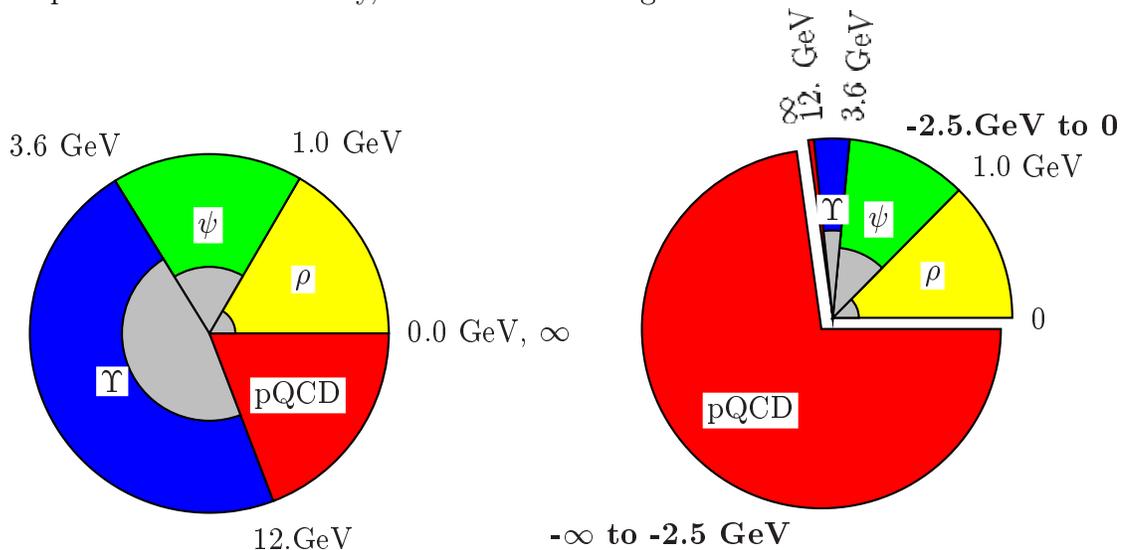}
\end{picture}

\caption{Comparison of the distribution of contributions and errors
(shaded areas scaled up by 10) in the standard (left) and the Adler
function based approach (right), respectively.}
\label{fig:alpsta}
\end{figure}

Our alternative procedure to evaluate $\dahzE$ in the Euclidean region
has several advantages as compared to other approaches used so far:
The virtues of our analysis are the following:
\begin{itemize}
\item no problems with the physical threshold and resonances
\item p-QCD is used only in the Euclidean region and not below 2.5 GeV.
 For lower scales p-QCD ceases to describe properly the functional
 dependence of the Adler function~\cite{EJKV98} (although the p-QCD answer
 remains within error bands down to about 1.6 GeV).
\item no manipulation of data must be applied and we need not refer to
 global or even local duality. That power corrections of the type
 Eq.~(\ref{NP}) are negligible has been known for a long time.  This,
 however, does not proof the absence of other kind of non-perturbative
 effects. Therefore our conservative choice of the minimum Euclidean
 energy seems to be necessary.
\item
 as we shall see our non--perturbative ``remainder'' $\dah0$ is mainly
 sensitive to low energy data, which changes the chances of
 possible future experimental improvement dramatically, as illustrated 
in Fig.~\ref{fig:alpsta}.
\end{itemize}

\vspace*{47mm}

The two methods (standard vs. Euclidean) of evaluating $\dahz$ are
also compared in Fig.~\ref{fig:alpsta}

\newpage

While the uncertainties to $\dahz$ in the standard approach are coming
essentially from everywhere below $M_\Upsilon$, which would make a
new scan over all energies for a precision measurement of
$\sigma_{\mathrm{had}}\equiv \sigma(e^+e^- \rightarrow \gamma^*
\rightarrow {\rm hadrons})$ unavoidable, the new approach leads to a
very different situation. The uncertainty of $\dah0$ is completely
dominated by the uncertainties of data below $M_{J/\psi}$ and thus
new data on $\sigma_{\mathrm{had}}$ are only needed below about 3.6 GeV
which could be covered by a tunable ``$\tau$--charm facility''.

Table~\ref{tab:alperr} compares our results with results obtained by
other authors which obtain smaller errors because they are using p-QCD
in a less controlled manner.

\begin{table}[ht]
\begin{tabular}{ll|c|c|cc}
\hline
$\dahz$ &$\delta \dal$ & $\delta \sinf$& $\delta M_W$ & Method & Ref. \\
\hline
0.0280~&0.00065~~ & 0.000232 & 12.0 & data $< ~12.~~\gv$                      &\cite{EJ95} \\
0.02777&0.00017~~ & 0.000061 &  3.2 & data $< ~1.8~~\gv$                      &\cite{KS98}\\
0.02763&0.00016~~ & 0.000057 &  3.0 & data $< ~1.8~~\gv$                      &\cite{DH98b}\\
0.027730&0.000209 & 0.000075 &  3.9 & Euclidean $> ~2.5~~\gv$                 &\cite{FJ98} \\
0.027426&0.000190  & 0.000070 &  3.6 & scaled data, pQCD 2.8-3.7, 5-$\infty$  &\cite{MOR00}\\
0.027649&0.000214  & 0.000078 &  4.0 & same but ``exclusive''                  &\cite{MOR00}\\
0.027896 & 0.000391  & 0.000139 &  7.2 &  ~\cite{EJ95} + new data
CMD \& BES      &\cite{FJ01} \\
~~~~~-&0.00007~~ & 0.000025 &  1.3 &{$\delta \sigma \:\lapprox\: 1\%$ up to $J/\psi$}\\
~~~~~-&0.00005~~ & 0.000018 &  0.9 &{$\delta \sigma \:\lapprox\: 1\%$ up to $\Upsilon$}\\
\multicolumn{2}{c|}{world average} & 0.000160 & 22.0 & Osaka 2000 \\
\hline
\end{tabular}
\caption{%
$\dahz$ and its uncertainties in different evaluations. Two
entries show what can be reached by increasing the precision of cross
section measurements to 1\%. $\delta M_W$ in MeV.}
\label{tab:alperr}
\end{table}

\section{The running electric charge at high energies}

Beyond the $Z$ peak not only the fermions contribute to the vacuum
polarisation but also the bosonic degrees of freedom in particular the
charged $W$--boson. However, if we try to ``define'' the running
charge in terms of the photon propagator simply, we get into
troubles. The analogy of a fermion loop, which is gauge invariant at
the one--loop level at least, the $W$--boson loop is not gauge
invariant. In fact one cannot measure self-energy contributions in
isolation. What experimentalists can measure are cross sections, in
the simplest case for a ``$2 \ra 2$ fermions'' process with
contributions from self--energy-, vertex- and box-diagrams. A
physically more acceptable definition of a running charge seems to be
via the electromagnetic form--factor of the electron, for example, but
also this is true only in an energy range where the one--photon
exchange approximation is accurate, such that we face a factorisation
of the cross section like in Thomson scattering at low energies. At
high energies (far off--shell) a form--factor is not any longer
accessible directly by experiment. We then may adapt a formal
definition, like the \MSb scheme, which is unphysical because heavy
degree of freedom do not decouple automatically in spite of the fact
that heavy states cannot affect the physics at much lower energies. In
the \MSb scheme one has to perform decoupling ``by hand'' therefore,
i.e., one only counts degrees of freedom which are lighter than a
given scale. That this may cause problems is not very surprising since
one tends to switch off individual members of gauge group multiplets.

Our analysis above, which includes non--perturbative effects from low
energy hadrons, is more in the spirit of on--shell renormalization,
which is more physical with respect to its decoupling behaviour. The
latter comes out for free in an on--shell scheme, because on--shell
renormalization exhibits the correct physical threshold
structure. But, as mentioned above, probing an on--shell electron by
an off--shell photon of virtuality $q^2$ is not physical, and in fact
not gauge invariant in the non--Abelian SM~\cite{JF85}. Still, a
reasonable {\em convention} is possible by requiring the photon to
satisfy Maxwell's equations, which is not automatic. The reason is
that in the SM, adopting the standard lowest order definition, the
photon field
\be
A_\mu =(gB_\mu+g'W_{\mu3})/\sqrt{g^{'2}+g^2}
\ee
has a non-Abelian component. This fact at higher orders causes
problems which do not appear in pure QED. A manifestly Abelian photon
may be defined by~\cite{JF85}
\ba
A^a_\mu &=&(gB_\mu+g'W^s_{\mu3})/\sqrt{g^{'2}+g^2} \crn
        &=&(\sqrt{g^{'2}+g^2}/g)\:B_\mu+(g'/g)\:Z^s_\mu
\ea
where $Z^s_\mu$ defined by
\ba
Z^s_\mu&=&(gW^s_{\mu3}-g'B_\mu)/\sqrt{g^{'2}+g^2}\crn
       &=&-(i/\sqrt{g^{'2}+g^2})\:(\Phi^+D_\mu \Phi-{\rm
h.c.})/(\Phi^+\Phi)
\ea 
obviously is a singlet, with respect to the SM gauge group, and
$W^s_{\mu 3}$ is Abelian. $\Phi$ is the Higgs doublet field and
$D_\mu\Phi$ its covariant derivative. In the {\em unitary gauge}
$A_\mu$ and $A^a_\mu$ coincide, which means that in the unitary gauge
we automatically are dealing with the Abelian photon field, which
satisfies the correct Maxwell equations. The gauge dependent part is
originating at the one--loop level solely from the $W$--pair
excitation, described by the diagrams\\

\hspace*{2.5cm} \epsfbox{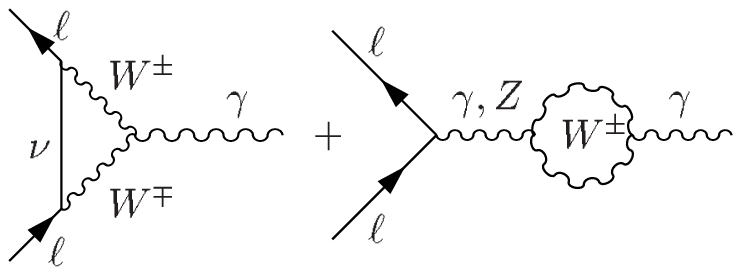}~~~.\\

Because we are mainly interested in the high energy behaviour and in
order to avoid lengthy expressions, we present the results for
$|q^2|\gg M_W^2$ only. The one--loop contributions to the singlet
form--factor may be written as
\be
\Delta \alpha = \Delta \alpha_{W} + \Delta \alpha_{Z} + \Delta
\alpha_{\gamma}+\Delta \alpha_{f}  
\ee
with contributions from $W$, $Z$, photon and the fermions.
For the ``renormalized'' virtual $W^\pm$ contribution one finds
\ba
\Delta \alpha_W(q^2) &=& \frac{\alpha}{16 \pi \stt \ctt} \Bigg\{
a_0+b_0 \ln \frac{|q^2|}{M_W^2}+
a_1\frac{q^2}{M_W^2} \left(\ln \frac{|q^2|}{\mu^2}-\frac83 \right)\crn 
&& ~~~~~~~~~~~~~~~~~~~~~~~~~~~~~
-i\pi \theta(q^2-4M_W^2)\left(b_0+a_1\frac{q^2}{M_W^2} \right)
\Bigg\} 
\ea
in the \MSb subtraction scheme. Introducing the notation $\ctt\doteq
M_W^2/M_Z^2,\:\stt\doteq 1-\ctt$ and $g(\ctt)\doteq\sqrt{4\ctt-1}~~
{\rm arc}~{\rm ctg} \sqrt{4\ctt-1}$, the coefficients are given by
\bea
a_0 &=& -32\cts -\frac{56}{3}\ctf+\frac{253}{6}\ctt +\frac12 \\
    &&  -(-32\cts
-\frac{64}{3}\ctf+\frac{134}{3}\ctt-\frac{22}{3}-\frac12 \ctit)g(\ctt)\\
b_0 &=& \frac53 \ctt -\frac{19}{6}-\frac14 \ctit\\
a_1 &=& \frac16 \ctt -\frac14\;\;.
\eea
The virtual $Z$ contribution reads
\ba
\Delta \alpha_Z(q^2) &=& \frac{\alpha}{16 \pi \stt \ctt} \Bigg\{
a'_0+b'_0 \ln \frac{|q^2|}{M_Z^2}+\frac23 b'_0\:\left({\rm
Sp}\left(1+\frac{q^2}{M_Z^2}\right)-\frac{\pi^2}{6}\right)\crn
&& ~~~~~~~~~~~~~~~~~~~~~~~~~~~~~
 -i\pi\theta(q^2-4m_e^2)\:b'_0\Bigg\} 
\ea
which in contrast to the $W^\pm$ contribution is finite (i.e.,
$\mu$--independent). The coefficients are given by
\bea
a'_0 &=& -14\ctf+21\ctt -\frac{35}{4}\\
b'_0 &=&  12\ctf-18\ctt +\frac{15}{2}\;\;,
\eea
and the Spence function Sp is asymptotically given by
\ba
{\rm Sp}\left(1+\frac{q^2}{M_Z^2}\right)-\frac{\pi^2}{6}\simeq
\left\{ \bary{lr} \frac{\pi^2}{6}-\frac12 \ln^2 \frac{q^2}{M_Z^2} +
i\pi \ln \frac{q^2}{M_Z^2}\:; & q^2 \gg M_Z^2~\\
-\frac{\pi^2}{3}-\frac12 \ln^2 \left(-\frac{q^2}{M_Z^2}\right)\:; 
& -q^2 \gg M_Z^2\:.
\eary \right.
\ea

The QED electron vertex $+$ self--energy contributions exhibit the
well known infrared problem with soft and collinear logs which only
become physical after combining them with the soft real photon
radiation. Virtual $+$ soft QED corrections together are related to
definition of the initial and/or final state and are therefore taken
into account in a different way. They have nothing to do with the
running of the charge or vacuum polarization effects. We therefore
apply the convention to set $\Delta_\gamma=0$ in the calculation of
$\dal$.

Another problem is due to $\gamma-Z$ mixing. At higher energies the
mixing effects have to be taken into account. We have seen that this
is crucial for the $W$--pair creation and reabsorption but in fact
also applies to the fermion contributions, once $\gamma \ra Z \ra
\gamma$ transitions become relevant at sufficiently high energies, we
must include
\bea
\Delta \alpha_f(q^2)=-2a \frac{\mz}{q^2-\mz}\:(A^{\gamma
Z}_{1r})_f-(ev)^2 \frac{1}{q^2}\:(A^{\gamma
\gamma}_{1r})_f
\eea
where
\bea
-(ev)^2 \frac{1}{q^2}\:(A^{\gamma
\gamma}_{1r})_f= \frac{\al}{3\pi} \sum_f Q_f^2 H_f\left( 4m_f^2/q^2\right)
\eea
is the renormalized QED vacuum polarization and 
\bea
(A^{\gamma Z}_{1r})_f= -\frac{\al}{24\pi \stt} \frac{q^2}{\mw}
\sum_f (4a_fQ_f)\left[ H_f\left( 4m_f^2/q^2\right)-
H_f\left( 4m_f^2/\mz\right)\right]
\eea
is the $\gamma-Z$ mixing contribution. The function $H_f$ is given by
\bea
H_f(y_f)= \frac{5}{3}+y_f+\left(1+ \frac{y_f}{2}\right)\:\sqrt{1-y_f}
\ln \frac{\sqrt{1-y_f}-1}{\sqrt{1-y_f}+1}
\eea
and $a_f=-Q_f\stt \pm \frac14$ for the upper and lower components of
the weak iso-doublets, respectively. The one--loop perturbative
fermion formula also is appropriate to take into account the {\bf top
quark} contribution. At $M_Z$ we have included in $\aiz$
\bea
\Delta \alpha_{\rm top}(\mz) = -0.76 \times 10^{-4}\;\;.
\eea
Numerical results for the SM contributions in the singlet form--factor
definition of the effective charge will be presented elsewhere.

\section{Concluding Remarks}

Experimental efforts to measure very precisely the total cross section
{$\sigma(\epm \ra hadrons)$} at low energies are mandatory for the
future of electroweak precision physics. Taking into account recent
theoretical progress, these ``low energy'' measurements are not only
important for testing the muon anomalous magnetic moment $\amu$ but as
well for the effective fine structure constant $\az$. A real
breakthrough would be possible by measuring $\sigma(\epm \ra hadrons)$
at 1\% accuracy below the $\tau$--threshold. We once more refer to
Tab.~\ref{tab:alperr} for the status and future perspectives.

Fortunately there is work in progress which can help to further reduce
the uncertainties of theoretical predictions: (i) VEPP-2M Novosibirsk
(CMD-2, SND): can further improve to 1.5\% up to 1.4 GeV. An upgrade
of the machine and the detectors is under consideration. (ii)
DA$\Phi$NE Frascati (KLOE): within one year of running we expect a
measurement below the $\phi$ resonance which is expected to be
competitive to the Novosibirsk data. Since the KLOE experiment is very
different in the technology from the Novosibirsk experiments this will
provide a very important cross check of older results. (iii) BEPC
Beijing (BES): can still improve in the important $J/\Psi$ region and
down to 2 GeV. (iv) In future a possible ``$\tau$--charm facility''
tunable between 2 GeV and 3.6 GeV would settle the remaining problems
essentially.\\[4mm]

If we adopt the Euclidean approach of calculating $\dahz$ via $\dah0$ ,
in future the Adler function is an ideal object for direct simulation
in lattice QCD. Of course, there is a long way to go, in order to
achieve an accuracy which is competitive with present evaluations from
the $\epm$--data (see Figs.~\ref{fig:Adlera} and ~\ref{fig:Adlerb}). However, a
purely theoretical prediction of $\Delta\alpha^{(5)}_{\rm
had}(-M_Z^2)$ seems to be feasible in future. Continuous progress in
theory and experiment let us expect that the necessary improvements
required for the future of precision physics will be realized. This is
particularly important for the electroweak precision physics which
would be possible with the Giga-Z--option at TESLA.

\section*{Acknowledgements}
It is a pleasure to thank Simon Eidelman, Jochem Fleischer, Oleg
Tarasov, Andrej Kataev and Oleg Veretin for fruitful collaboration on
the topics presented in this note.

\bb{99}

\bibitem{TDR} TESLA, Technical Design Report, DESY-2001-011, ECFA 2001-209.

\bibitem{LEP} The LEP Collaborations ALEPH, DELPHI, L3, OPAL and ...,
hep-ex/0101027 (unpublished).

\bibitem{Sirlin80} A. Sirlin, \PRD 22 (1980) 971.

\bibitem{Veltman77} M.~Veltman, Nucl. Phys. {\bf B123} (1977) 89.

\bibitem{TopTev95} F.~Abe et al., Phys. Rev. Lett. {\bf 74} (1995) 2626; 
S.~Abachi et al., Phys. Rev. Lett. {\bf74} (1995) 2632.

\bibitem{KalSab55}
G. K\"all\'en and A. Sabry, {\it K. Dan. Vidensk. Selsk. Mat.-Fys. Medd.}
{\bf 29} (1955) No. 17.

\bibitem{Ste98}
M. Steinhauser, {\it Phys. Lett.} {\bf B429} (1998) 158.

\bibitem{GKL}
S.G. Gorishny, A.L. Kataev and S.A. Larin, {\it Phys. Lett.} {\bf B 259} (1991)
144;\\
L.R. Surguladze and M.A. Samuel, {\it Phys. Rev. Lett.} {\bf 66}
   (1991) 560; ibid. 2416 (Err),\\
K.G. Chetyrkin, {\it Phys. Lett.} {\bf B 391} (1997) 402.

\bibitem{ChK95}
 K.G. Chetyrkin and J.H. K\"uhn, \PLB {\bf 342} (1995) 356 and references therein.

\bibitem{ChHK00}
K.G. Chetyrkin, R.V. Harlander, J.H. K\"uhn, hep-ph/0005139.

\bibitem{EJ95}
S.I.~Eidelman and F.~Jegerlehner, {\it Z. Phys.} {\bf C 67} (1995) 585;\\
F.~Jegerlehner, {\it Nucl. Phys. B} (Proc. Suppl.) {\bf 51C} (1996) 131.

\bibitem{CMD}
R.R. Akhmetshin et al., Preprint BINP 99-10, Novosibirsk, 1999.

\bibitem{BES}
J.Z. Bai et al.(BES Collaboration), hep-ex/0102003.

\bibitem{BP95} H. Burkhardt, B. Pietrzyk, \PLB {\bf 356}, 398 (1995)

\bibitem{ADH98} R. Alemany, M. Davier, A. H\"ocker,
                {\it Eur.~Phys.~J.} {\bf C 2}, 123 (1998)

\bibitem{MOR00}
A.D. Martin, J. Outhwaite, M.G. Ryskin, \PLB {\bf 492} (2000) 69,
hep-ph/0012231.

\bibitem{BP01}
H. Burkhardt, B. Pietrzyk, Preprint LAPP-EXP-2001-03. 

\bibitem{PDG}
D.E. Groom et al..(Particle Data Group), {\it Eur. Phys. J.} {\bf C 15} (2000)
1.
 
\bibitem{MZ95} A.D. Martin, D. Zeppenfeld, \PLB {\bf 345}, 558 (1995)

\bibitem{DH98a} M. Davier, A. H\"ocker, \PLB {\bf 419}, 419 (1998)

\bibitem{KS98} J.H. K\"uhn, M. Steinhauser, \PLB {437}, 425 (1998)

\bibitem{GKNS98} S. Groote, J.G. K\"orner, N.F. Nasrallah, K. Schilcher,
                 \PLB {\bf 440}, 375 (1998)

\bibitem{DH98b} M. Davier, A. H\"ocker, \PLB {\bf 435}, 427 (1998)

\bibitem{Erler98} J. Erler, hep-ph/9803453

\bibitem{FJ86} F. Jegerlehner, \ZPC {\bf 32}, 195 (1986)

\bibitem{FJ01} F. Jegerlehner, LC-TH-2001-035 (this note).

\bibitem{SVZ} M.A. Shifman, A.I. Vainshtein and V.I. Zakharov,
              \NPB {\bf 147}, 385 (1979)

\bibitem{mqcd3} K.G. Chetyrkin, J.H. K\"uhn, M. Steinhauser,
               \PLB {\bf 371}, 93 (1996);
               \NPB {\bf 482}, 213 (1996); B {\bf 505}, 40 (1997); 
               K.G.~Chetyrkin, R.~Harlander, J.H.~K\"uhn, M.~Steinhauser,
               \NPB {\bf 503}, 339 (1997)

\bibitem{JT98}
F.~Jegerlehner and O.V.~Tarasov, \NPB {\bf 549} (1999) 481.

\bibitem{FJ98} F. Jegerlehner, in ``Radiative Corrections'',
ed.~J. Sol\`a, World Scientific, Singapore, 1999

\bibitem{EJKV98} S.~Eidelman, F.~Jegerlehner, A.L.~Kataev, O.~Veretin,
Phys. Lett. B454 (1999) 369.

\bibitem{JF85} F.~Jegerlehner, J.~Fleischer, {\it Phys. Lett.} {\bf 151B} 
(1985) 65, {\it Acta Phys. Pol.} {\bf B17} (1986) 709.

%
%

\end{thebibliography}

\end{document}